\documentclass[12pt]{iopart}
\usepackage{iopams,graphicx}
\begin{document}

\title{Constraints on the lepton asymmetry and radiation energy density: Implications for {\sc Planck}}

\author{L.A. Popa and  A.Vasile}{
\address{ISS Institute for Space Sciences Bucharest-Magurele, Ro-077125 Romania}

\eads{\mailto{lpopa@venus.nipne.ro},
\mailto{avasile@venus.nipne.ro}}

\date{\today}

\begin{abstract}

By using  most of the present Cosmic Microwave Background (CMB) and  Large Scale Structure (LSS)
measurements and the Big Bang Nucleosynthesis (BBN) constraints
on the primordial helium abundance, $Y_p$,
we set bounds on the radiation content of the Universe and neutrino properties.
We consider lepton asymmetric cosmological models parametrized by the neutrino degeneracy
parameter $\xi_{\nu}$ and the variation of the relativistic degrees of freedom,
$\Delta  N_{oth}^{eff}$, due to possible other physical processes that occurred between BBN
and structure formation epoch.\\
We found that present CMB and LSS data constraints the neutrino degeneracy
parameter at $\xi_{\nu} \leq 0.722$,
implying a lepton asymmetry of the neutrino background $ {\cal L}_{\nu} \leq 0.614$ (2-$\sigma$).
We also found  $\Delta N^{eff}_{oth}=0.572 ^{+1.972}_{-1.780}$ ,
the contribution to the effective number of relativistic neutrino species
$ N^{eff}=3.058^{+1.971}_{-1.178}$ and a primordial helium abundance
$Y_p=0.249^{+0.014}_{-0.016}$ (2-$\sigma$ errors).\\
These results bring an important improvement over the similar ones obtained by using WMAP~1-year and older LSS data or
the WMAP~3-year data alone and the standard primordial helium abundance value $Y_p=0.24$,
relaxing the stringent BBN constraint on the neutrino degeneracy parameter ($\xi_{\nu} \leq 0.07$).

We forecast that the CMB temperature and polarization maps observed with high angular
resolutions and sensitivity by the future {\sc Planck} Mission will
constraint the primordial  primordial helium abundance at $Y_p=0.247 \pm 0.002$
(2-$\sigma$ errors) in agreement with the most stringent limits on $Y_p$ given
by the BBN  and the neutrino degeneracy parameter at $\xi_{\nu} \leq 0.280$ (2-$\sigma$),
not excluding the possibility of larger lepton asymmetry.\\

This work has been done on behalf of {\sc Planck}-LFI activities.

\end{abstract}
\pacs{CMBR theory, dark matter, cosmological neutrinos, gravitational lensing }

\maketitle

\section{Introduction}

The radiation budget of the Universe relies on a strong theoretical prejudice:
apart from the Cosmic Microwave Background (CMB) photons, the relativistic background
would consist of neutrinos and of possible contributions from other relativistic relicts.
The main constraints on the radiation energy density come either from the very early Universe, where the radiation was the dominant source of energy, or from the observation of cosmological perturbations which
carry the information about the time equality between matter and radiation.

\vspace{0.3cm}
In particular, the primordial light element abundance predictions in the standard theory
of the Big Bang Nucleosynthesis (BBN) \cite{Wagoner,Olive,Burles,Eidelman}
depend on the baryon-to-photon ratio, $\eta_B$, and the
radiation energy density at the BBN epoch (energy density of order MeV$^4$), usually parametrized by the effective number of relativistic neutrino species, $N^{eff}$.

Meanwhile, the number of active neutrino flavors have been fixed by  $Z^0$ boson decay
width to $N_{\nu}=2.944 \pm 0.012$ \cite{Eidelman} and the combined study of the incomplete neutrino decoupling and the QED corrections indicate that the number of relativistic neutrino
species is $N^{eff}_{\nu}=3.046$ \cite{Mangano02}.  Any departure of  $N^{eff}$ from this last value would be due to non-standard neutrino features or to the contribution of other relativistic relics. \\
The solar and atmospheric neutrino oscillation experiments  \cite{Fuk98,Amb98} indicate the existence of non-zero neutrino masses in eV range.

There are also indications
of neutrino oscillations with  larger mass-squared difference,
coming from the short base-line oscillation experiments \cite{Atha,Miniboone},
 that can be explained by adding one or two sterile neutrinos with eV-scale mass to the standard scheme with three active neutrino flavors (see Ref.\cite{Maltoni} for a recent analysis). Such results have impact on cosmology because sterile neutrinos can contribute to the number of relativistic degrees of freedom at the Big Bang Nucleosynthesis \cite{Cirelli}.
These models are subject to strong bounds on the sum of active neutrino masses from the combination of various cosmological data sets \cite{Rafelt,Find},
ruling out a thermalized sterile neutrino component with eV mass \cite{Seljak07,Mel}.\\
However, there is the possibility to accommodate the cosmological observations
with data from short base-line neutrino oscillation experiments by postulating
the existence of a sterile neutrino with the mass of few keV having a phase-space distribution significantly suppressed relative to the thermal distribution.\\
For both, non-resonant zero lepton number production and enhanced resonant production
with initial cosmological lepton number, keV sterile neutrinos are produced via small mixing angle oscillation conversion of thermal active neutrinos \cite{Aba02}.\\
Sterile neutrino with mass of few keV provides also a
valuable Dark Matter (DM) candidate \cite{DodWidr,Dol,Aba01,Sha06},
alleviating the accumulating contradiction between the $\Lambda$CDM model
predictions on small scales and observations, by smearing out the small scale structure.\\
On the other hand, the possible existence of new particles such as axions and gravitons, the time variation of the physical constants and  other non-standard scenarios (see e.g. \cite{Sarkar96} and references therein) could contribute to the radiation energy density at BBN epoch.

At the same time, more phenomenological extensions to the standard neutrino sector
have been studied, the most natural being consideration of the leptonic asymmetry
\cite{Freese83,Ruffini83,Ruffini88}, parametrized by the neutrino degeneracy parameter $\xi_{\nu}=\mu_{\nu}/T_{\nu_0}$
[$\mu_{\nu}$ is the neutrino chemical potential and
$T_{\nu_0}$  is the present temperature of the neutrino background,
$T_{\nu_0}/T_{\rm cmb}=(4/11)^{1/3}$]. \\
Although  the standard model predicts the leptonic asymmetry  of the same order
as the  baryonic asymmetry, $B \sim 10^{-10}$, there are many particle physics scenario
in which a leptonic asymmetry much larger can be generated \cite{Smith06,Cirelli06}.
One of the cosmological implications of a larger leptonic asymmetry is
the possibility to generate small baryonic asymmetry of the Universe through
the non-perturbative (sphaleron) processes \cite{Kuzmin85,Falcone01,Buch04}.
Therefore, distinguishing between a vanishing and non-vanishing $\xi_{\nu}$ at the BBN epoch is a crucial test of the standard assumption that sphaleron effects equilibrate the
cosmic lepton and baryon asymmetries.\\
The measured neutrino mixing parameters implies that neutrinos
reach the chemical equilibrium before BBN \cite{Dolgov02,Wong02,Beacom02} so that  all
neutrino flavors are characterized by the same
degeneracy parameter, $\xi_{\nu}$, at this epoch.\\
The most important impact of the leptonic asymmetry on BBN is
the shift of the beta equilibrium between protons and neutrons and the increase
of the radiation energy density parametrized by:
\begin{equation}
\Delta N^{ eff}(\xi_{\nu})=3\left[ \frac{30}{7}
\left(\frac{\xi_{\nu}}{\pi} \right)^2
+\frac{15}{7}\left( \frac{\xi_{\nu}}{\pi} \right)^4 \right] \,.
\end{equation}
The BBN constraints on N$^{eff}$ have been recently reanalyzed
by comparing  the theoretical predictions and experimental data on the
primordial abundances of light elements,
using the baryon abundance derived from  the
WMAP~3-year (WMAP3) CMB temperature and  polarization measurements
\cite{Spergel07,Nolta,Page}: $\eta_B=6.14 \times 10^{-10}(1.00\pm 0.04)$.
In particular, the $^4$He abundance, $Y_p$,  is quite sensitive to the value of $N^{eff}$.\\
The analysis of Ref. \cite{Mangano07}, the conservative error analysis
of  helium abundance,
$Y_P=0.249 \pm 0.009$ \cite{Olive04}, yielded to $N^{eff}=3.1^{+1.4}_{-1.2}$ (2-$\sigma$) in good agreement with the standard value,
but still leaving some room for non-standard values, while
more stringent error bars of helium abundance , $Y_p=0.2516 \pm 0.0011$ \cite{Izotov07}, leaded to $N^{eff}=3.32^{+0.23}_{-0.24}$ (2-$\sigma$) \cite{Ichikawa07}.

The stronger constraints on the degeneracy parameter obtained from BBN  \cite{Serpico05} gives $-0.04 < \xi< 0.07$ (1-$\sigma$), adopting  the conservative error analysis of $Y_p$ of Ref. \cite{Olive04} and $\xi=0.024 \pm 0.0092$ (1-$\sigma$),
adopting the more stringent error bars of $Y_p$ of Ref. \cite{Izotov03}.

\vspace{0.3cm}
The CMB anisotropies and LSS matter density fluctuations power spectra carry the signature
of the energy density of the Universe at the time of matter-radiation equality
(energy density of order eV$^4$), making possible the measurement of $N^{eff}$ through its effects on the
growth of cosmological perturbations. \\
More effective number of relativistic neutrino species enhances the integrated Sachs-Wolfe effect on the CMB power spectrum, leading to a higher first acoustic Doppler peak amplitude.
Also, the delay of the epoch of matter-radiation equality shifts the LSS matter power spectrum turnover position toward larger angular scales, suppressing the power at small scales. In particular, for the leptonic asymmetric models, the neutrino mass is lighter than in the symmetric case. This leads to changes in neutrino free-streaming length and neutrino Jeans mass due to the increase of the neutrino velocity dispersion \cite{Lattanzi05,Ichiki07}.

After WMAP3 data release, there are many works aiming to constrain $N^{eff}$ from cosmological observations \cite{Seljak07,Spergel07,Mangano07,Han06,Cirelli06b,Kawa07}.
Their results suggest large values for $N^{eff}$ within 2-$\sigma$ interval, some of them  not including the standard value $N^{eff}_{\nu}=3.046$ \cite{Seljak07,Spergel07,Mangano07}. Recently Ref. \cite{Han07} argues that the discrepancies are due to the treatment of the  scale-dependent biasing in the galaxy power spectrum inferred from the main galaxy sample of the Sloan Digital Sky Survey data release 2 (SDSS-DR2)  \cite{Teg04a,Teg04b}
and the large fluctuation amplitude reconstructed from  the Lyman$-\alpha$ forest data
\cite{McDonald05} relative to that inferred from WMAP3. \\
Discrepancies between  BBN  and cosmological data results on $N^{eff}$ was interpreted as  2-$\sigma$ evidence of the fact that further relativistic species are produced by particles decay between BBN and structure formation \cite{Cirelli06b,Kawa07}.
Other theoretical scenarios include the violation of the spin-statistics in the neutrino sector \cite{Dolgov05}, the possibility of an extra interaction between
the dark energy and radiation or dark matter, the existence of a Brans-Dicke field which could mimic the effect of adding extra relativistic energy density between BBN and structure formation epochs \cite{deFelice}.

\vspace{0.3cm}
The extra energy density can be split in two distinct uncorrelated contributions, first due to net lepton asymmetry of the neutrino background and second due to the extra contributions from other unknown processes:
\begin{equation}
\Delta N^{eff}=\Delta N^{eff}(\xi)+ \Delta N_{oth}^{eff}\, .
\end{equation}
The aim of this paper is to obtain bounds on the neutrino lepton asymmetry and on the extra radiation energy density by using most of the existing CMB and LSS measurements and self-consistent BBN priors on $Y_p$.
We also to compute the sensitivity of the future {\sc Planck} experiment \cite{Blue} for these parameters testing
the restrictions on cosmological models with extra relativistic degrees of freedom expected from high precision CMB temperature and polarization anisotropy measurements.     \\

The paper is organized as follows: Section~2 contains a review on the lepton asymmetric cosmological models and the BBN theory, Section~3 is devoted to a summary data analysis method while in Section~4 we discuss our results.
We draw our main conclusions in Section~5

\section{Leptonic asymmetric cosmological models and the BBN theory}

\begin{figure}
\begin{center}
\includegraphics[height=10cm,width=10cm]{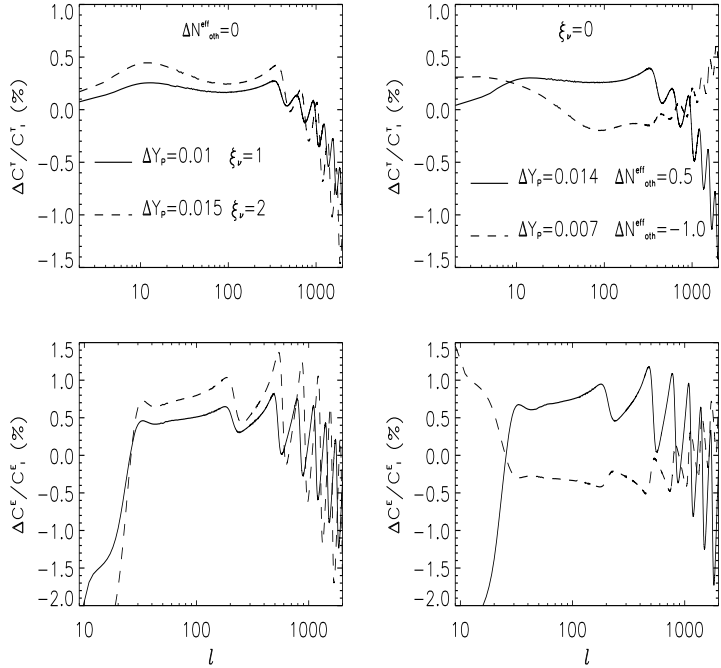}
\caption{The CMB temperature (top panels) and polarization (bottom panels)
percentage differences corresponding to different helium fractions
variations $\Delta Y_p$ with respect to the standard value $Y_p=0.248$ for:
$ \xi \neq 0$ and $\Delta N^{eff}_{oth}=0$~(left panels) and
$\Delta N^{eff}_{oth} \neq 0$ and $ \xi =0$~(right panels).
All other parameters are fixed to the values of our fiducial model.}
\label{fig1}
\end{center}
\end{figure}
The density perturbations in leptonic asymmetric cosmological models
have been discussed in literature \cite{Lattanzi05,Ichiki07,Pastor99,Orito02}.
We applied them to modify the Boltzmann Code for Anisotropies in the Microwave Background (CAMB) \cite{Hu00,Cha,Lewis} to compute the  CMB temperature and polarization anisotropies
power spectra  and LSS matter density fluctuations power spectra for the case of
three degenerate neutrinos/antineutrinos with mass $m_{\nu}$ and degeneracy parameter $\xi_{\nu}$. As neutrinos reach their approximate chemical potential
equilibrium before BBN epoch \cite{Dolgov02,Wong02,Beacom02}, we consider in our computation that all three flavors of neutrinos/antineutrinos have the same degeneracy parameter $\xi_{\nu}$. For simplicity, we also consider all there neutrino/antineutrino flavors with  the same mass $m_{\nu}$. \\
When the Universe was hot enough, neutrinos and antineutrinos of each flavor
behave like relativistic particles with Fermi-Dirac phase space distributions:
\begin{eqnarray}
f_{\nu}(q)=\frac{1}{e^{E_{\nu}/T_{\nu}-\xi_{\nu}}+1} \, ,
\hspace{0.2cm}
f_{{\bar \nu}}(q)=\frac{1}{e^{E_{{\bar \nu}}/T_{\nu}-\xi_{{\bar \nu}}}+1} \,,
\end{eqnarray}
where $E_{\nu}=\sqrt{q^2+a^2m_{\nu}}$ is one flavor neutrino/antineutrino energy
and $q=ap$ is the comoving momentum.  Hereafter, $a$ is the cosmological scale factor ($a_0=1$ today). The mean energy density and pressure of one flavor of massive degenerated neutrinos and antineutrinos can be written as:
\begin{eqnarray}
\rho_{\nu}+\rho_{\bar \nu}=(k_BT_{\nu})^4\int^{\infty}_0 \frac{d^3q}{(2\pi)^3} \, q^2 E_{\nu}(f_{\nu}(q)+f_{\bar \nu}(q))\, ,\\
3(P_{\nu}+P_{\bar \nu})=(k_BT_{\nu})^4\int^{\infty}_0 \frac{d^3q}{(2\pi)^3} \, \frac{q^2} {E_{\nu}}(f_{\nu}(q)+f_{\bar \nu}(q))\, .
\end{eqnarray}
We modify in CAMB the expressions for the energy density and the pressure in the relativistic and non-relativistic limits for the degenerate case \cite{Ichiki07}
and  follow the standard procedure to compute the perturbed quantities  by expanding the phase space distribution function of neutrinos and antineutrinos into homogeneous  and perturbed inhomogeneous components \cite{Lewis,Ma95,Seljak96}.
Since the gravitational source term in the Boltzmann equation is proportional to the logarithmic derivative of the neutrino distribution function with respect to comoving momentum, ${\rm d} \ln (f_{\nu}+f_{\bar \nu})/ {\rm d} \ln q$, we also modify this term
to account for $\xi_{\nu} \neq 0$ \cite{Ichiki07,Pastor99}.

As mentioned in the first section, the BBN theory gives strong constraints on $N^{eff}$
at this epoch by comparing the measured light element abundance with the theoretical predictions. The only free parameter is the baryon to photon ratio, $\eta_B=n_b/n_{\gamma}$, that is obtained from the CMB observation of $\Omega_b h^2$.\\
In particular, the $^{4}He$ mass fraction,$Y_p$, affects the CMB angular power spectra through its impact on different evolution phases of the ionization/recombination history   \cite{Trota03}. \\
We modify the recombination routine {\bf recfast} \cite{Scott99} of the CAMB code to explicitly account for the dependence of $Y_p$ on $\Omega_b h^2$ and on $\Delta N^{eff}$
as defined in equation (2), as previously suggested in Ref. \cite{Iki_He},  by adopting the fitting formula \cite{Serpico04}:
\begin{eqnarray}
\hspace{-2.5cm}10Y_p= \left(\sum^8_{n=1}  a_n x^{n-1} +
\sum^8_{n=1} b_n x^{n-1} \Delta N^{eff} +
\sum^8_{n=1} c_n x^{n-1} (\Delta N^{eff})^2+
\sum^8_{n=1} d_n x^{n-1} (\Delta N^{eff})^3 \right)  \nonumber \\
\hspace{-1cm}\times \, {\rm exp} \left( \sum^6_{n=1} e_n x^{n} \right)\, , &
\end{eqnarray}
where $x={\rm log}_{10}(10^{10}) \eta$ and $10^{10} \eta=273.49 \Omega_b h^2$.
The coefficients $a_n$, $b_n$, $c_n$, $d_n$ and $e_n$
are given in Ref. \cite{Serpico04}. The standard prediction of BBN $Y_p=0.248$ is obtained
for $\Delta N^{eff}=0$. According to Ref. \cite{Serpico04}, the accuracy of this fitting formula is better than 0.05\% for the range
$5.48 \times 10^{-10} < \eta_B <  7.12 \times 10^{-10}$
($0.02 < \Omega_b h^2 < 0.026$)
which corresponds to the 3-$\sigma$ interval obtained by WMAP3 and
$-3 < \Delta N^{eff} < 3$.\\
Figure~1 presents the CMB temperature and polarization
percentage differences corresponding to the different
variations of the helium fraction, $\Delta Y_p$, with respect to
the standard value $Y_p=0.248$ obtained for
$ \xi \neq 0$ and $\Delta N^{eff}_{oth}=0$ and
$\Delta N^{eff}_{oth} \neq 0$ and $ \xi =0$.
The impact of the percent change in $Y_p$ on the
ionization/recombination history has a net impact on the CMB
temperature and polarization power spectra at percent level.

\section{Analysis}

We use the {\sc CosmoMC}  Monte Carlo Markov Chain (MCMC) public package \cite{Bridle}
modified for our extended $6+3$ parameter space to sample from the posterior distribution giving the following experimental datasets.\\
{The Cosmic Microwave Background} (CMB): We  use the WMAP3 data \cite{Spergel07,Nolta,Page} complemented with  the CMB data from Boomerang \cite{Netterfield02,Tavish05}, ACBAR \cite{Kuo02} and CBI \cite{Readhead04} experiments.\\
{Large Scale Structure} (LSS):
The power spectrum of the matter density fluctuations has been inferred from
the galaxy clustering data of the Sloan digital Sky Survey (SDSS) \cite{Teg04a,Teg04b,Tegmark06,Percival07} and Two-degree Field Galaxy Redshift Survey (2dFGRS) \cite{Cole05}. \\
In particular, the luminous red galaxies (LRG) sample from the SDSS
data release 5 (SDSS-DR5)  has more statistical significance  \cite{Tegmark06,Percival07}
than the spectrum retrieved from the SDSS main galaxy sample from data release 2 (SDSS-DR2)
\cite{Teg04a,Teg04b}  eliminating the existing tension between the power spectra
from SDSS-DR2 and 2dFGRS.
For this reason we consider in our analysis the matter power spectra from SDSS-LRG and 2dFGRS. We consider SDSS-LRG data up to $k_{max} \simeq 0.2$h Mpc$^{-1}$ and the 2dFGRS data up to $k_{max} \simeq 0.14$h Mpc$^{-1}$. We apply the corrections due to the non-linearity behavior and scale dependent bias as indicated in Ref.\cite{Tegmark06}, connecting the linear matter power spectrum, $P_{lin}(k)$,  and the galaxy power spectrum, $P_{gal}(k)$, by:
\begin{equation}
P_{gal}(k)=b^2 \frac{1+Q_{nl}k^2}{1+1.4k}P_{lin}(k) \,,
\end{equation}
where the free parameters  $b$ and $Q_{nl}$ are marginalized.\\
{Type Ia Supernovae} (SNIa):
We also use the luminosity distance measurements of distant Type Ia supernovae
obtained by Supernova Legacy Survey (SNLS) \cite{Astier06}
and the Hubble Space Telescope \cite{Riess04}.\\
{Hubble Space Telescope key project} (HST):
We impose  priors on the Hubble constant
$H_0=72\pm 8$ km s$^{-1}$Mpc$^{-1}$ from HST key project \cite{Freedman01}.\\
{BBN constraints on $Y_p$}:
We use the BBN constraints on $Y_p$ as given in equation (7), allowing $\Omega_b h^2$
and $\Delta N^{eff}$ to span the following ranges:
$0.02 < \Omega_b h^2 < 0.026$ and $-3 < \Delta N^{eff} < 3$. \\
Hereafter, we will denote  WMAP3+SDSS-DR5+2dFGRS+SNIa+HST+BBN data set as WMAP3+All.

\vspace{0.5cm}
We perform our analysis in the framework of the extended
$\Lambda$CDM cosmological model described by $6+3$ free parameters:
\begin{eqnarray}
\Theta=( \underbrace{\Omega_b h^2,\Omega_{cdm} h^2, \theta_s, \tau, n_s, A_s,}_{standard} f_{\nu},
\xi_{\nu}, \Delta N_{oth}^{eff} )\, .
\end{eqnarray}
Here $\Omega_{b} h^2$ and $\Omega_{cdm}h^2$ are the baryonic and cold dark matter energy density parameters, $\theta_s$ is the ratio of the sound horizon distance to the angular diameter distance,
$\tau$ is the reionization optical depth, $n_s$ is the scalar spectral index of the primordial density perturbation power spectrum and $A_s$ is its amplitude at the pivot scale $k_*=0.05$~hMpc$^{-1}$. The additional three parameters denote the neutrino energy density fraction $f_{\nu}$, the neutrino degeneracy parameter $\xi_{\nu}$ and the extra contributions from other
unknown processes $\Delta N_{oth}^{eff}$. Table~1 presents
the  parameters of our model, their fiducial values used
to generate the {\sc Planck}-like simulated power spectra and the prior ranges
adopted in the analysis.

\begin{table}
\begin{center}
\caption{The free parameters of our model, their fiducial values used
to generate the {\sc Planck}-like simulated power spectra and the prior ranges
adopted in the analysis.}
\end{center}
\vspace{0.3cm}
\begin{center}
\begin{tabular}{lll}
\hline \hline
 Parameter & Fiducial value  & Prior range   \\
\hline \hline
$\Omega_{b}h^2$ & 0.022 & 0.005 $\rightarrow$ 0.1 \\
$\Omega_{cdm}h^2$ & 0.105& 0.01 $\rightarrow$ 0.5 \\
$\theta_s$ &1.04 & 0.5 $\rightarrow$ 5 \\
$\tau$& 0.09& 0.01 $\rightarrow$ 0.3  \\
$n_s$ & 0.95 & 0.5 $\rightarrow$ 1.3 \\
${\rm ln}[10^{10}A_s]$ & 3 & 2.7 $\rightarrow$ 4 \\
$f_{\nu}$ & 0.05 & 0 $\rightarrow$ 0.5 \\
$\xi_{\nu}$ & 0 & 0 $\rightarrow$ 4 \\
$\Delta N^{neff}_{oth}$& 0.046 & -3 $\rightarrow$ 3 \\
$Y_p$ & 0.248 & \\
\hline \hline
\end{tabular}
\end{center}
\end{table}
For the forecast from {\sc Planck}-like simulated data we use the CMB temperature (T) and
polarization (P) power spectra of our fiducial cosmological model and	
the expected  experimental characteristics of the {\sc Planck} frequency channels
presented in Table~2 \cite{Blue}. For each frequency channel we consider an homogeneous
detector noise with the power spectrum:
\begin{eqnarray}
N^{c}_{l,\nu}=(\theta_b \Delta_a)^2 \exp^{ l(l+1) \theta^2_b / 8 \ln 2}
\hspace{0.3cm} c \in(T,P) \,,
\end{eqnarray}
where $\nu$ is the frequency of the channel, $\theta_b$ is the FWHM of the  beam and
$\Delta_c$ are the corresponding sensitivities per pixel.
The global noise of the experiment is obtained as:
\begin{equation}
N^c_l=\left[ \sum_{\nu} (N^{c}_{l,\nu})^{-1} \right]^{-1} \, .
\end{equation}

\begin{table}
\begin{center}
\caption{The expected  experimental characteristics
for the {\sc Planck} frequency channels considered in the paper.
$\Delta_T$ and $\Delta_P$ are the sensitivities per pixel for temperature
and polarization maps.}
\end{center}
\vspace{0.3cm}
\begin{center}
\begin{tabular}{cccc}
\hline \hline
 $\nu $ & FWHM  & $\Delta_T  $ & $\Delta_P $ \\
 (GHz)&(arc-minutes)&    ($\mu$ K)&       ($\mu$ K)                           \\
\hline \hline
               100 & 9.5 &6.8 & 10.9 \\
               143 & 7.1 &6.0 & 11.4 \\
               217 & 5.0& 13.1 & 26.7 \\
\hline \hline
\end{tabular}
\end{center}
\end{table}
In order to interpret the likelihood function, ${\cal L}({ \Theta})$,
as probability density  we assume uniform prior probability
on the parameters $\Theta$ (i.e. will assume that all values of parameters are equally probable).
For each parameter we compute the cumulative distribution function
$C(\theta)=\int^{\Theta}_{\Theta_{min}} {\cal L }
(\Theta)d \Theta/ \int_{\Theta_{min}}^{\Theta_{max}} {\cal L }d \Theta$
and quote as upper and lower intervals at 2-$\sigma$ the values at which $C(\theta)$ is 0.95 and 0.05 respectively.  For the case when  ${\cal L}({ \Theta})$ is zero and $\Theta$ has a positive values (i.e. $f_{\nu}$, the absolute value of $\xi_{\nu}$)
we quote only the upper limit at 2-$\sigma$.\\

\section{Results}

We start by making a consistency check, verifying that by using WMAP3+All data
and imposing  $\xi_{\nu}=0$, $\Delta N^{eff}_{oth}=0$ and
$Y_p=0.248$ priors we obtain results in agreement with the ones
obtained by WMAP collaboration (Tables 5 and 6 from Ref. \cite{Spergel07}).

In order to understand how the extra relativistic energy density
and the leptonic asymmetry affect the determination of  other
cosmological parameters, we compute first the likelihood functions  for WMAP3+All
by imposing $\Delta N^{eff}_{oth}=0$ prior.
We then extend our computation over the whole parameter space for WMAP3+All
and {\sc Planck}-like simulated data.
\begin{figure}
\begin{center}
\includegraphics[height=8cm,width=8cm]{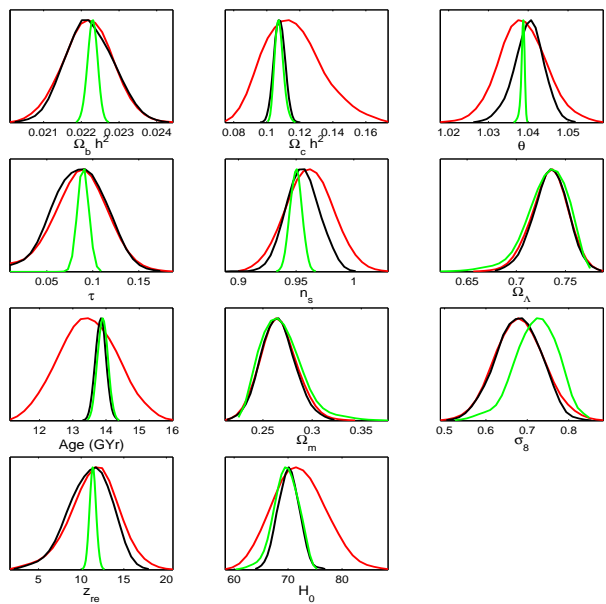}
\caption{The marginalized posterior likelihood probabilities of the main cosmological
parameters obtained for: WMAP3+All with $\Delta N^{eff}_{oth}=0$ prior (black lines) and
WMAP3+All (red lines) and {\sc Planck} (green lines) without priors on  $\Delta N^{eff}_{oth}$.}
\label{fig1}
\end{center}
\end{figure}
In Figure~2  we compare the marginalized likelihood  probabilities
obtained for WMAP3+All with $\Delta N^{eff}_{oth}=0$ prior with those obtained for
WMAP3+All and {\sc Planck} without priors on  $\Delta N^{eff}_{oth}$.
The main effect of including the contribution of the extra relativistic energy density
is the change in the age of the Universe (and in the Hubble expansion rate)
from $t_0=13.81 \pm 0.26$ GYrs to  $t_0=13.42^{+1.3}_{-1.42}$ GYrs (2-$\sigma$ errors),
effect  that is mostly driven by the increased degeneracy
between matter and radiation energy densities. \\
We  present in Figure~3 the 2D marginalized 1-$\sigma$ and 2-$\sigma$
allowed regions in $t_0$ -  $|\xi|$ and  $t_0$ -  $\Delta N^{eff}_{oth}$ planes showing
this effect.
\begin{figure}
\begin{center}
\includegraphics[height=5.33cm,width=8.66cm]{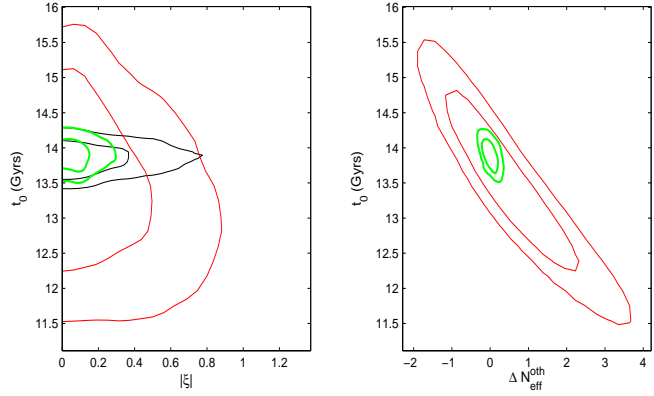}
\caption{The 2D marginalized 1-$\sigma$ and 2-$\sigma$
allowed regions in $t_0$ -  $|\xi|$ and  $t_0$ -  $\Delta N^{eff}_{oth}$ planes.
The black curves correspond to
WMAP3+All with $\Delta N^{eff}_{oth}=0$ prior.
The red and green curves correspond respectively
to WMAP3+All and {\sc Planck}-like simulated data,
without priors on  $\Delta N^{eff}_{oth}$.}
\label{fig1}
\end{center}
\end{figure}
The 2-$\sigma$ confidence region  for the additional number of relativistic relicts
is $ -1.207 \leq \Delta N^{eff}_{oth} \leq 2.572 $ for WMAP3+All and
 $ -0.226 \leq \Delta N^{eff}_{oth} \leq 0.236 $ for  {\sc Planck}-like simulated data.
The negative values of $\Delta N^{eff}_{oth}$ are lowering
the amplitude of the CMB and LSS power spectra that can be
compensated by larger values of the degeneracy parameter.\\
Figure~4 presents the 2D marginalized 1-$\sigma$ and 2-$\sigma$
allowed regions in $\sigma_8$ - $n_s$ plane
and  in $n_s$ -  $\tau$ plane
obtained for WMAP3+All with  $\Delta N^{eff}_{oth}=0$ prior and
WMAP3+All  and {\sc Planck}-like simulated data  without
priors on $\Delta N^{eff}_{oth}$. It is evident from these plots that
the inclusion of an  additional number of relativistic relicts
increases the degeneracy between the cosmological parameters controlling the
CMB and LSS power spectra amplitudes.
\begin{figure}
\begin{center}
\includegraphics[height=5.33cm,width=8.66cm]{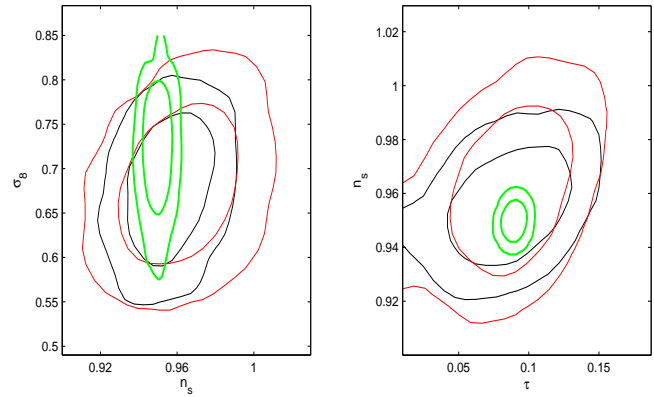}
\caption{The 2D marginalized 1-$\sigma$ and 2-$\sigma$
allowed regions in $\sigma_8$ - $n_s$ plane (left panel)
and the same confidence regions in $n_s$ -  $\tau$ plane (right panel)
obtained for WMAP3+All with  $\Delta N^{eff}_{oth}=0$ prior (black lines) and
WMAP3+All (red lines) and {\sc Planck}-like simulated data (green lines) without
priors on $\Delta N^{eff}_{oth}$.}
\label{fig1}
\end{center}
\end{figure}

\vspace{0.2cm}
In Figure~5  we compare the marginalized likelihood  probabilities
of the neutrino parameters and the helium mass fraction
obtained for WMAP3+All with $\Delta N^{eff}_{oth}=0$ prior with those obtained for
WMAP3+All and {\sc Planck} without priors on  $\Delta N^{eff}_{oth}$.
The expectation values and the corresponding  errors or the upper limits (2-$\sigma$)
are presented in Table~3.\\
\begin{figure}
\begin{center}
\includegraphics[height=8.cm,width=8.66cm]{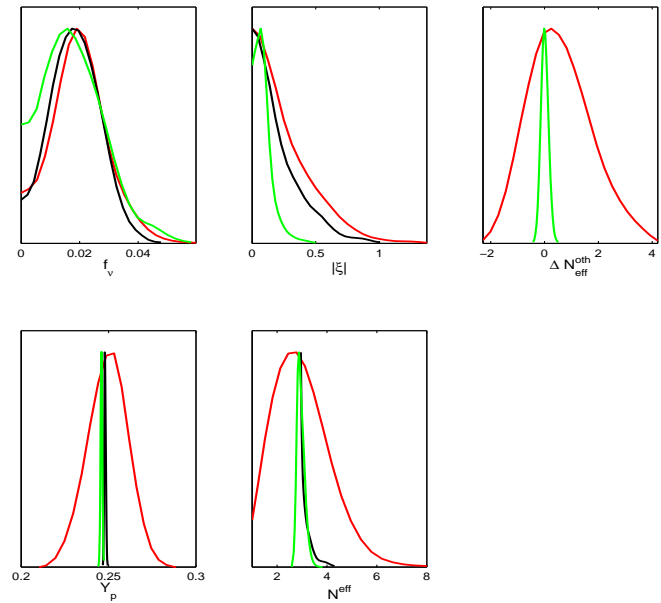}
\caption{The marginalized posterior likelihood probabilities of the neutrino parameters
and helium mass fraction obtained for: WMAP3+All with $\Delta N^{eff}_{oth}=0$ prior (black lines) and  WMAP3+All (red lines) and {\sc Planck} (green lines) without priors on  $\Delta N^{eff}_{oth}$.}
\label{fig1}
\end{center}
\end{figure}
\begin{table}
\begin{center}
\caption{Constraints on neutrino parameters and helium mass fraction.
The errors and the upper limits are at 2-$\sigma$.}
\vspace{0.3cm}
\begin{tabular}{llll}
\hline
          & WMAP3+All & WMAP3+All& {\sc Planck} \\
Parameter & $\Delta N^{eff}_{oth}=0$ & $\Delta N^{eff}_{oth} \ne 0$ &  $\Delta N^{eff}_{oth} \ne 0$  \\
\hline \hline
$f_{\nu}$              &  $\leq 0.033  $     & $\leq 0.037  $    & $\leq 0.036 $ \\
$|\xi|$                  &  $ \leq 0.590 $     & $ \leq 0.722 $    & $\leq 0.280$ \\
$\Delta N^{eff}(\xi)$  &  $ \leq 0.833  $    & $\leq 1.243  $    & $\leq 0.158$  \\
${\cal L}_{\nu}$       &  $ \leq 0.474 $     & $ \leq 0.614 $   & $ \leq 0.179$  \\
$\Delta N^{eff}_{oth}$ &  $ - $                &  $0.572 ^{+1.972}_{-1.780}$ &
$0.008^{+0.229}_{-0.234}$ \\
$N^{eff}$       &  $\leq 3.873$       &  $ 3.058^{+1.971}_{-1.178}$   &
$2.920^{+0.267}_{-0.216} $ \\
$Y_p$ & $\leq 0.249$   & $0.249^{+0.014}_{-0.016}$ & $0.247 \pm 0.002$\\
\hline
\end{tabular}
\end{center}
\end{table}
\begin{figure}
\begin{center}
\includegraphics[height=5.33cm,width=8.66cm]{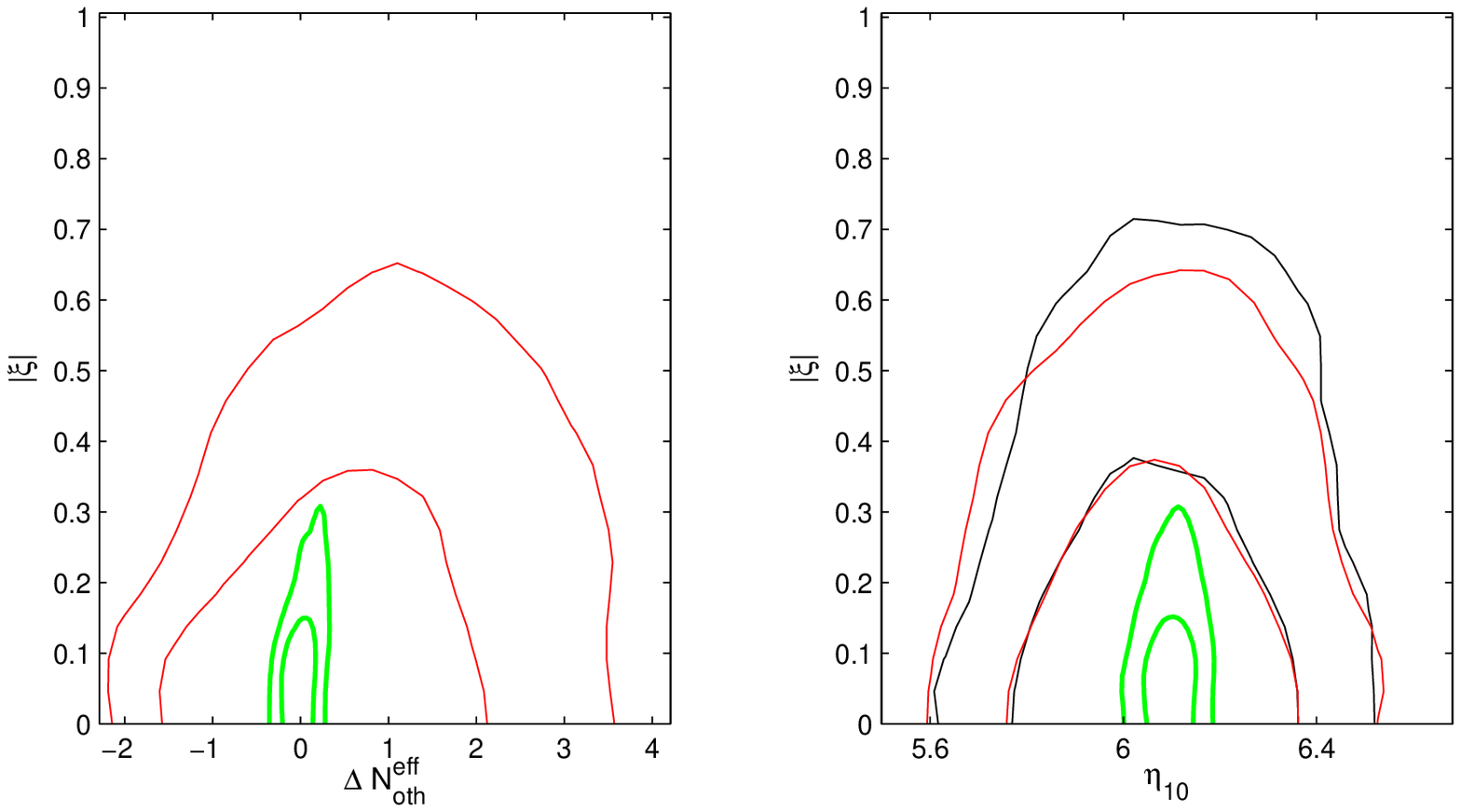}
\caption{The 2D marginalized 1-$\sigma$ and 2-$\sigma$
allowed regions in $|\xi|$ - $\Delta N^{eff}_{oth}$ and \\ $|\xi|$ - $\eta_{10}$ planes.
The black curves correspond to WMAP3+All with $\Delta N^{eff}_{oth}=0$ prior.
The red and green curves correspond respectively to WMAP3+All and {\sc Planck}-like simulated data,
without priors on  $\Delta N^{eff}_{oth}$.}
\label{fig1}
\end{center}
\end{figure}
We show in Figure~6 that the precise measurements of the CMB temperature and polarization power spectra
from {\sc Planck} will reduce the degeneracy between  $|\xi|$ and $\Delta N^{eff}_{oth}$,
allowing  better constrains of the scenarios involving additional number of
relativistic relicts and leptonic asymmetry.\\

\newpage

\section{Conclusions}

In this paper, we set bounds on the radiation content of the Universe and neutrino properties by using most of the present Cosmic Microwave Background (CMB) and  Large Scale Structure (LSS) measurements.
We also take into account the Big Bang Nucleosynthesis (BBN) constraints on the primordial helium abundance, $Y_p$, which prove to be important in the estimation of cosmological parameters from future {\sc Planck} data, both in non-degenerate and degenerate BBN models; the importance of the self-consistent BBN prior on $Y_p$ was also emphasized in two other recent analysis \cite{Hamann07, Ichikawa071}.\\
We consider lepton asymmetric cosmological models parametrized by the neutrino degeneracy
parameter $\xi_{\nu}$ and the variation of the relativistic degrees of freedom,
$\Delta  N_{oth}^{eff}$, due to possible other physical processes that occurred between BBN
and structure formation epoch.\\
We found that present CMB and LSS data together with BBN prior on the primordial helium abundance ($Y_p$) constraints the neutrino degeneracy
parameter at $\xi_{\nu} \leq 0.722$, leading to a lepton asymmetric neutrino background of $ {\cal L}_{\nu} \leq 0.614$ (2-$\sigma$).
We also found  $\Delta N^{eff}_{oth}=0.572 ^{+1.972}_{-1.780}$ , the contribution to the effective number of re-\\lativistic neutrino species
$ N^{eff}=3.058^{+1.971}_{-1.178}$ and a primordial helium abundance $Y_p=0.249^{+0.014}_{-0.016}$ (2-$\sigma$ errors).
These values represent important improvements over the similar results obtained by using WMAP~1-year together with older LSS data \cite{Crotty03} or
the WMAP3 data alone \cite{Lattanzi05} and the standard primordial helium abundance value $Y_p=0.24$,
relaxing the stringent BBN constraint on the neutrino degeneracy parameter ($\xi_{\nu} \leq 0.07$).

We observe that, when using WMAP3+All data, an additional number of relativistic relicts brings a substantial degeneracy in the $\Omega_{m} - \Omega_{r}$ plane and a weaker constraint on the age of the Universe; the same degeneracy occurs also between other cosmological parameters under the same conditions. We therefore conclude that the present cosmological data do not favor the variation of the relativistic degrees of freedom, $\Delta N_{oth}^{eff}$, due to other possible physical processes that occurred between BBN and matter-radiation equality epoch.\\

We forecast that the CMB temperature and polarization maps observed with high angular resolutions and sensitivity by the future {\sc Planck} Mission will
constraint the primordial  primordial helium abundance at $Y_p=0.247 \pm 0.002$ (2-$\sigma$ errors) in agreement with the most stringent limits on $Y_p$ given by the BBN  and the neutrino degeneracy parameter at $\xi_{\nu} \leq 0.280$ (2-$\sigma$), allowing larger lepton asymmetry models.
Also, they will reduce the degeneracy between $|\xi|$ and $\Delta N_{oth}^{eff}$ allowing a better distinction between extra radiation energy density coming from an additional number of relativistic relicts and from a lepton asymmetric neutrino background.\\

{\bf Acknowledgements}
L.P. and A.V. acknowledge the support by the ESA/PECS Contract "Scientific exploitation of  Planck-LFI data"

We also acknowledge the use of the GRID computing system facility at the Institute for
Space Sciences Bucharest and would like to thank the staff working there.







\section*{References}

\begin{thebibliography}{99}

\bibitem{Wagoner}
Wagoner R V, Fowler W A, Hoyle F, 1967
{\it Astrophys. J.} {\bf 148} 3

\bibitem{Olive} Olive K A, Steigman G, Walker, T P, 2000
{\it Phys. Rep.} {\bf 333} 389 [astro-ph/9905320]

\bibitem{Burles} Burles S,  Nollett K  M, Turner M S, 2001
{\it Astrophys. J.} {\bf 552} L1 [astro-ph/0010171]

\bibitem{Eidelman} Eidelman S et al, 2004
{\it Phys. Lett.} {\bf B 592} 1

\bibitem{Mangano02} Mangano G,  Miele G , Pastor S, Peloso M, 2002
{\it Phys. Lett. } {\bf B 534} 8 [astro-ph/0111408]

\bibitem{Fuk98} Fukuda Y et al (Super-Kamiokande Collab.), 1998
{\it Phys. Rev. Lett.} {\bf 81} 1562

\bibitem{Amb98} Ambrosio M et al (MACRO Collab.), 1998
{\it Phys. Lett.} {\bf B 434} 451 [hep-ex/9807005]

\bibitem{Atha} Athanassopoulos C et al, 1996
{\it Phys. Rev. Lett.} {\bf 77} 3082 [nucl-ex/9605003]

\bibitem{Miniboone} Aguilar-Arevalo A A et al (MiniBooNE Collaboration) 2007
[hep-ex/0704.1500]

\bibitem{Maltoni} Maltoni M. ans Schwets T, 2007
[hep-ex/0705.0107]

\bibitem{Cirelli} Cirelli M , Marandella G, Strumia S, Vissani F, 2005
{\it Nucl. Phys.} {\bf B 708}  215 [hep-ph/0403158]

\bibitem{Rafelt} Hannestad S and  Raffelt G G, 2006
{\it J. Cosmol.  Astropart. Phys.} JCAP11(2006)016 [astro-ph/0607101]

\bibitem{Find} Kristiansen J,  Eriksen H K and Elgar A, 2006
{\it Phys. Rev.} {\bf D 74} 123005 [astro-ph/0608017]

\bibitem{Seljak07} Seljak U, Solsar A and McDonald P, 2006
{\it J. Cosmol. Astropart. Phys.} JCAP10(2006)014 [astro-ph/0604335]

\bibitem{Mel}  Dodelson S, Melchiorri A and  Slosar A, 2006
{\it Phys. Rev. Lett.} {\bf 97} 041301 [astro-ph/0511500]

\bibitem{Aba02} Abazajian K N and Fuller G M, 2002
{\it Phys. Rev.} {\bf D 66} 023526 [astro-ph/0204293]

\bibitem{DodWidr} Dodelson S, Widrow L M, 1994
{\it Phys. Rev. Lett.} {\bf 72}, 17 [hep-ph/9303287]

\bibitem{Dol} Dolgov A D  and Hansen S H, 2002
{\it Astropart. Phys.} {\bf 16} 339 [hep-ph/0009083]

\bibitem{Aba01} Abazajian K, Fuller G M, Patel M, 2001
{\it Phys. Rev.} {\bf D 64} 023501 [astro-ph/0101524]

\bibitem{Sha06} Asaka T, Sahaposhnikov M and Kusenko A, 2006
{\it Phys. Lett.}  {\bf B 638} 401 [hep-ph/0602150]

\bibitem{Sarkar96} Sarkar S,  1996
{\it Rept. Prog. Phys} {\bf 59} 1493  [hep-ph/9602260]

\bibitem{Freese83} Freese K, Kolb E W, Turner M S, 1983
{\it Phys. Rev. } {\bf D27} 1689

\bibitem{Ruffini83} Ruffini R,  Song D J,  Stella L, 1983
{\it Astron. Astrophys.} {\bf 125} 265

\bibitem{Ruffini88} Ruffini R, Song D J, Taraglio S, 1988
{\it Astron. Astrophys.} {\bf 190} 1


\bibitem{Smith06} Smith C  J, Fuller G M,  Kishimoto C T, Abazajian K N, 2006
{\it Phys. Rev.} {\bf D 74} 085008 [astro-ph/0608377]

\bibitem{Cirelli06} Chu Y Z, Cirelli M, 2006
{\it Phys. Rev.} {\bf D 74} 085015

\bibitem{Kuzmin85} Kuzmin V, Rubakov V, Shaposhnikov M, 1985
{\it Phys. Lett.} {\bf B 155} 36

\bibitem{Falcone01} Falcone D, Tramontano F, 2001
{\it Phys. Rev.} {\bf D 64} 077302 [hep-ph/0102136]

\bibitem{Buch04} Buchmuller W,  Di Bari P, Plumacher M, 2004
{\it New Jour. Phys.} {\bf 6} 105 [hep-ph/0406014]

\bibitem {Dolgov02} Dolgov A D, Hansen S H, Pastor S, Petcov S T, Raffelt G, Semikoz D V, 2002 {\it Nucl.Phys.} {\bf B 632}, 363 [hep-ph/0201287]

\bibitem{Wong02} Wong Y Y Y, 2002
{\it Phys. Rev.} {\bf D 66} 025015 [hep-ph/0203180]

\bibitem{Beacom02} Abazajian K N,  Beacom J F, Bell N F, 2002
{\it Phys. Rev.} {\bf D 66} 013008 [astro-ph/0203442]

\bibitem{Spergel07} Spergel D N et al.(wmap cOLLABORATION), 2007
{\it Astrophys. J. Suppl.} {\bf 170} 377 [astro-ph/0603449]

\bibitem{Nolta} Hinshaw et al. (WMAP Collaboration),2007
{\it Astrophys. J. Suppl.} {\bf 170} 288  [astro-ph/0603451]

\bibitem{Page} Page L et al. (WMAP Collaboration), 2007
{\it Astrophys. J.Suppl.} {\bf 170} 335 [astro-ph/0603450]

\bibitem{Mangano07} Mangano G, Melchiorri A, Mena O, Miele G, Slosar A, 2007
{\it J. Cosmol.  Astropart. Phys.} JCAP03(2007)006 [astro-ph/0612150]

\bibitem{Olive04} Olive K A,  Skillman E D, 2004
{\it Astrophys. J.} {\bf 617} 29 [astro-ph/0405588]

\bibitem{Izotov07} Izotov Y I, Thuan T X,  Stasinska G, 2007
{\it Astrophys. J.} {\bf 662} 15 [astro-ph/0702072]

\bibitem{Ichikawa07} Ichikawa K, Kawasaki M, Nakayama K, Senami M,  Takahashi F, 2007
{\it J. Cosmol.  Astropart. Phys.} JCAP05(2007)008 [hep-ph/0703034]

\bibitem{Serpico05} Serpico P D, Raffelt G G, 2005
{\it Phys.Rev. } {\bf D 71} 127301 [astro-ph/0506162]

\bibitem{Izotov03} Izotov Y I,  Thuan T X, 2004
{\it Astrophys. J.} {\bf 602} 200 [astro-ph/0310421]

\bibitem{Lattanzi05} Lattanzi M, Ruffini R, Vereshchagin G V,
{\it Phys. Rev.} {\bf D 72} 063003  [astro-ph/0509079]

\bibitem{Ichiki07} Ichiki k , Yamaguchi M,  Yokoyama J, 2007
{\it Phys.Rev} {\bf D 75}, 084017 [hep-ph/0611121]

\bibitem{Han06} Hannestad S,  Raffelt G G, 2006
{\it J. Cosmol. Astropart. Phys.} JCAP11(2006)016 [astro-ph/0607101]

\bibitem{Cirelli06b} Cirelli M, Strumia A, 2006
{\it J. Cosmol. Astropart. Phys.} JCAP12(2006)013 [astro-ph/0607086]

\bibitem{Kawa07} Ichikawa K, Kawasaki M, Takahashi F, 2007
{\it J. Cosmol. Astropart. Phys.} JCAP05(2007)007 [astro-ph/0611784]

\bibitem{Han07} Hamann J, Hannestad S, Raffelt G G, Wong Y Y Y, 2007
[arXiv:0705.0440)]

\bibitem{Teg04a} Tegmark M et al (SDSS Collaboration), 2004
{Astrophys. J.} {\bf 606} 702 [astro-ph/0310725]

\bibitem{Teg04b} Tegmark M et al (SDSS Collaboration), 2004
{\it Phys. Rev.} {\bf  D 69} 103501  [astro-ph/0310723]

\bibitem{McDonald05} McDonald P et al, 2005
{\it Astrophysical J.} {\bf 635} 761 [astro-ph/0407377]

\bibitem{Dolgov05} Dolgov A D, Hansen S H, Smirnov A Y, 2005
{\it J. Cosmol. Astropart. Phys.} JCAP06(2005)004 [astro-ph/0611784]

\bibitem{deFelice} de Felice A,  Mangano G,  Serpico P D, Trodden M, 2006
{\it Phys. Rev.} {\bf D 74} 103005  [astro-ph/0510359]

\bibitem{Blue} The {\sc Planck} Consortia, 2005
"The Scientific Programme of {\sc Planck}"
{\it ESA-SCI} 1 [astro-ph/0604069 ]

\bibitem{Pastor99} Lesgourgues L, Pastor S, 1999
{\it Phys. Rev.} {\bf D 60} 103521 [hep-ph/9904411]

\bibitem{Orito02} Orito M, Kajino T, Mathews G J, Wang Y, 2002
{\it Phys. Rev.} {\bf D 65} 123504 [astro-ph/0203352]

\bibitem{Hu00} Hu W, 2000
{\it Phys.Rev.}  {\bf D 62} 043007  [astro-ph/0002238]

\bibitem{Cha} Challinor A, Lewis A, 2005
{\it Phys. Rev.}, {\bf D 71} 103010 [astro-ph/0502425]

\bibitem{Lewis} Lewis A, Challinor A, Lasenby A, 2000
{\it Astrophys. J.} {\bf 538} 473 [astro-ph/9911177]
\footnote{http://camb.info}

\bibitem {Ma95} Ma C P, Bertschinger E, 1995
{\it Astrophys. J.} {\bf 455} 7 [astro-ph/9401007]

\bibitem {Seljak96} Seljak U,  Zaldarriaga M, 1996
{\it Astrophys. J.} {\bf 469} 437  [astro-ph/9603033]

\bibitem{Trota03} Trotta R, Hansen S H, 2004
{\it Phys. Rev.} {\bf D 69} 023509 [astro-ph/0306588]

\bibitem{Scott99} Seager S, Sasselov D D, Scott D, 1999
{Astrophys. J.} {\bf 523}  L1-L5 [astro-ph/9909275]

\bibitem{Iki_He} Ichikawa K, Takahashi T, 2006
{\it Phys. Rev.} {\bf D 73} 063528 [astro-ph/0601099]

\bibitem{Serpico04} Serpico P D, Esposito S, Iocco F, Mangano G,
Miele G, Pisanti O, 2004
{\it J. Cosmol. Astropart. Phys.} JCAP12(2004)010 [astro-ph/0408076]

\bibitem{Bridle}  Lewis A and Bridle S, 2002
{\it Phys. Rev.}  {\bf D 66} 103511 [astro-ph/0205436]
\footnote{http://cosmologist.info/cosmomc}

\bibitem {Netterfield02} Netterfield C B et al.,2002
{\it Astrophys. J.} {\bf 571}  604  [astro-ph/0104460]

\bibitem {Tavish05} MacTavish C J et al., 2006
{\it Astrophys. J.} {\bf 647} 799 [astro-ph/0507503]

\bibitem {Kuo02}  Kuo C I et. al., 2004
{\it Astrophys. J.} {\bf 600} 32 [astro-ph/0212289]

\bibitem {Readhead04} Readhed A C S et al., 2004
{\it Astrophys. J.} {\bf 609} 498 [astro-ph/0402359]

\bibitem {Tegmark06} Tegmark M et al.(SDSS Collaboration), 2006
{\it Phys. Rev.} {\bf 74} 123507 [astro-ph/0608632]

\bibitem{Percival07} Percival W J et al., 2007
{\it Astrophys. J.}  {\bf 657} 645 [astro-ph/0608636]

\bibitem {Cole05} Cole S et al.(2dFGRS Collaboration), 2002
{\it Mon. Not.  R. Astron. Soc.} {\bf 362} 505 [astro-ph/0501174]

\bibitem{Astier06} Astier P et al., 2006
{\it Astron. Astophys.} {\bf 447} 31 [astro-ph/0510447]

\bibitem {Riess04} Riess A G et al., 2004
{\it Astrophys. J.} {\bf 607} 665 [astro-ph/0402512]

\bibitem{Freedman01} Freedman W L  et al., 2001
{\it Astrophys. J.} {\bf 553} 47 [astro-ph/0012376]

\bibitem{Hamann07} Hamann J, Lesgourgues J, Mangano G, 2007, astro-ph/0712.2826

\bibitem{Ichikawa071} Ichikawa K, Sekiguchi T, Takahashi T, astro-ph/0712.4327

\bibitem{Crotty03} Crotty P, Lesgourgues J, Pastor S, 2004
{\it Phys. Rev} {\bf D 69} 123007 [astro-ph/0302337]
\end{thebibliography}
\end{document}